\documentstyle[preprint,aps,psfig]{revtex}
\begin{document}
\def\be{\begin{equation}}
\def\ee{\end{equation}}
\title{A real-space, real-time method for the dielectric function}
\author{G.F. Bertsch$^{(a)}$\footnote{E:mail: bertsch@phys.washington.edu},
 J.-I. Iwata$^{(b)}$,
        Angel Rubio$^{(c)}$\footnote{E:mail: arubio@mileto.fam.cie.uva.es },
        and K. Yabana$^{(b)}$\footnote{E:mail: yabana@nucl.ph.tsukuba.ac.jp}}
\address{
$^{(a)}$Department of Physics and National Institute for Nuclear Theory,\\
University of Washington, Seattle, WA USA\\
$^{(b)}$Institute of Physics, University of Tsukuba,\\
        Tsukuba 305-8571, Japan\\
$^{(c)}$ Departamento de F\'{\i}sica Te\'orica, Universidad de Valladolid,
Valladolid\\ and Donostia International Physics
Center, San Sebastian, Spain\\
}
\maketitle
\begin{abstract}
We present an algorithm to calculate the linear response 
of periodic systems in the time-dependent density functional theory,
using a real-space representation of the electron wave functions 
and calculating the dynamics in real time.  The real-space formulation
increases the efficiency for calculating the interaction, and the
real-time treatment decreases storage requirements and allows the
entire frequency-dependent dielectric function to be calculated at
once. We give as examples the dielectric functions of a simple metal,
lithium, and an elemental insulator, diamond.
\end{abstract}
\section{Introduction}
     Real-space methods have proven their utility in calculating the
linear response of finite systems in time-dependent density functional
theory\cite{va99,ya96}.  However, there has 
been the perception that real-space methods are unsuitable for infinite 
periodic systems. The problem is that the long range polarization 
currents are important
but are dynamically independent of the local state of the electrons
within the unit cell.  Stated differently, the polarization gives
rise to a surface charge at the surface of any finite sample, but
the resulting electric field is independent of the charge density within
any cell in the interior.  The necessity to introduce the polarization
as an independent degree of freedom has been well recognized 
in the literature of the density functional theory\cite{re94,go95,or98,re99}.

    We will show here that in fact it is straightforward to treat 
infinite systems in the real-time formulation of TDDFT, simply by
adding as one additional dynamic variable the surface charge.  Formally,
this is conveniently done adding a gauge field in within a Lagrangian.
A gauge formalism has also been very recently applied 
by Kootskra et al \cite{ko00}, however using a frequency
representation rather than solving real-time equations.  In our formulation,
we derive the 
dynamic equations from the Lagrangian:
\be
 L = \int_{\Omega}d^3r\Bigg({\sum_i |\vec\nabla \phi_i/i- 
e A {\hat z}\phi_i|^2\over 2 m } + 
{1\over 8\pi }\vec\nabla V(r)\cdot \vec\nabla V(r) + e n(r)V(r)
+e n_{ion}(r)
V(r)  +
\ee
$$
{\cal
V}_{xc}[n(r)] +
{\cal V}_{ion}[\rho(r,r')]\Bigg)
-{\Omega\over 8\pi } \left({d A \over d t}\right)^2-i\int_\Omega d^3 r 
\sum_i\phi_i^*{\partial \phi_i\over \partial t}.
$$
Here the $\phi_i$ are the Bloch wave functions of the electrons, 
normalized so that $n(r) = \sum_i |\phi_i(r)|^2$ is the electron
density.  The volume of the unit cell is $\Omega$.
The electromagnetic interaction 
is separated into a Coulomb field
$V({\vec r})$ that satisfies periodic boundary conditions in the unit
cell and
a vector gauge field ${\hat z}A(t)$.  Note that the gauge field is uniform, without any
dependence on $r$.  The electric field is then given by
$$
{\vec{\cal E}} = -{\vec \nabla} V -{\hat z} {d A \over d t}.
$$
In these formal equations, we use units with $\hbar=c=1$.

The other pieces of the first integral are the usual terms in the 
Kohn-Sham energy functional. The term $e n(r) V(r) $ gives the
direct Coulomb interaction of the electrons, except for the
surface charging.  The ionic interaction 
is separated\footnote{The separation is somewhat
arbitrary, but is useful because the periodicity of $V$ then takes
the ionic
lattice into account automatically.}  into a long-range part that can be 
associated with  an ionic charge density $n_{ion}(r)$ and a short-range part
${\cal V}_{ion}$. The latter depends on the
orbital angular momentum of the electrons in typical {\it ab initio} 
pseudopotentials.  It therefore depends on the full one-electron 
density matrix $\rho(r,r') = \sum_i \phi^*_i(r) \phi_i(r')$.
We have
emphasized this point because nonlocal interactions do not respect gauge
invariance.  The invariance is of course restored if the density 
matrix is gauged.  If one expresses the nonlocality in terms of the operator
$\nabla$, one makes the replacement $\nabla\rightarrow \nabla-ie A {\hat
z}$. Finally, the 
${\cal V}_{xc}$ is the usual
exchange-correlation energy density of density functional theory.

Requiring the Lagrangian action to 
be stationary gives equations of motion for $\phi_i$ and $A$ and
the Poisson equation for $V$.  The dynamic equation for $\phi_i$
is the time-dependent Kohn-Sham equation,
\be
\label{tdlda}
-{\nabla^2\over 2 m}\phi_i  -{e \over m i} A \nabla_z \phi_i
+
{e^2\over 2 m} A^2 \phi_i + (e V +{\delta{\cal V}_{ion} \over 
\delta n} + {\delta{\cal V}_{xc}\over \delta n}
 )\phi_i
= i{\partial\over\partial t}
\phi_i
\ee
The equation for $A$ is
\be
{\Omega\over 4 \pi}{d^2 A\over d t^2}-
{e\over m}\sum_i\langle \phi_i|\nabla_z/i |\phi_i\rangle 
+{e^2\over m} A N_e+ {\delta\over \delta A}
\int_\Omega {\cal V}_{ion} d^3r =0
\ee
where
$N_e = \int_\Omega d^3r n(r)$
is the number of electrons per unit cell.  

\section{Linear Response, Sum Rule and Simple Models}
The calculation of the dielectric function using the above real-time
dynamic equations is very similar to the corresponding calculation of
dynamic polarizability of finite systems\cite{ya96}.  We first solve the
static equations (with $A$=0) to get the ground state electron orbitals
and the periodic Coulomb potential $V$.  The system is then perturbed by
making a sudden change in $A$, $A(t=0_+) = A_0$.  This corresponds to 
applying a short-duration electric field at $t=0$, ${\cal E}(t) = - A_0
\delta(t)$.  The dynamic equations are then applied to evolve the variables
in time, finding the time evolution of the polarization electric field
 ${\cal E}(t) = - d A (t)
/d t$.  The dielectric function $\epsilon (\omega)$ is just the
ratio of the Fourier components of the external and the total fields;
it is given by
\be
\label{epsilon}
{1\over \epsilon (\omega)} -1 = {1\over A_0} \int_{0_+}^{\infty} e^{i\omega
t-\eta t}
{d A(t)\over d t} dt.
\ee
Here $\eta$ is a small quantity to establish the imaginary part of
the response.  In principle the resulting theory automatically respects
the Kramers-Kronig relation.   

The linear energy-weighted sum rule is easily derived in this 
formalism.  The sum rule may be expressed as
\be
\int_0^\infty \omega{\rm Im\,\,} \epsilon^{-1}( \omega) d\omega
= - {2 \pi^2 e^2 N_e\over m \Omega}.
\ee
To calculate the sum rule with our Lagrangian, we write the 
integral using eq. (4)
\be
\int_0^\infty \omega{\rm Im\,\,} \epsilon^{-1}( \omega) d\omega
= {1\over A_0} \int_0^\infty dt {d A \over d t} {\rm Im}\int_0^\infty
d \omega \omega e^{i\omega t -\eta t}
={\pi \over 2 A_0}   \left( {d^2 A \over d t^2} \right)_{t=0_+}.
\ee
The second derivative in the last expression can be easily found
from eq. (3).  At $t=0_+$, the wave functions have not yet had
time to change, $A(0_+)=A_0$ and $\langle \nabla_z \rangle =0$.  
Then if the last term in eq. (3) can be neglected, 
\be
{d^2 A\over d t^2} = - {4 \pi e^2 N_e A_0 \over m\Omega}
\ee
and eq.~(5) follows immediately.  Thus the time-dependent treatment
satisfies the sum rules automatically to the extent permitted by
the last term.  That term is only nonzero for nonlocal pseudopotentials,
and in fact it may improve the accuracy of the theory by incorporating
effects of the core electrons on the dynamic properties\cite{footnote}.

Let us now see how the gauge field treatment works in a simple
analytically solvable model, namely the electron gas.
As mentioned before, when the field $A_0$ is applied, there is no 
immediate response
to the operator $\nabla$, since the wave function does not change 
instantaneously. However, in the Fermi gas, the single-particle states are 
eigenstates of momentum  so the response remains $\langle \nabla\rangle = 0$ for 
all time.  Putting this in
eq. (3), and dropping the pseudopotential term, the equation for
$A$ becomes simple harmonic motion, with the solution
\be
   A(t) = A_0 \cos \omega_{pl} t
\ee
where $\omega_{pl}$ is the plasmon frequency,
\be
  \omega_{pl}^2 = { 4 \pi e^2 N_e \over m \Omega} =  {4 \pi e^2 n \over
m}.
\ee
The dielectric function may now be calculated from the time integral
eq. (4). One obtains the familiar electron gas result,
\be
\epsilon(\omega) = 1 -{\omega_{pl}^2\over\omega^2}.
\ee
One sees that the derivation here is much simpler than the usual one
using the Coulomb gauge.  There one 
formulates the response in a particle-hole representation, and takes the
external field to be of the form $e^{iq\cdot r}$ with 
$q$ finite.  The dielectric function is then found by taking the
$q\rightarrow 0$ 
limit.

We can make another simple model for the opposite extreme of a tightly
bound electron in the unit cell.  Assume that the ion potential $e V_{ion}(r)+
{\delta{\cal V}_{ion} / 
\delta n} $ can be approximated by a harmonic oscillator potential
in the region over which the electron wave function is appreciable.  
According to 
Kohn's theorem\cite{do94}, the response is just the same as for 
an isolated electron in the same ionic potential.  This comes out of 
eq. (2-3) in the following way.  The initial impulse $A_0$ starts the 
electron 
moving, and as a result both $V(r)$ and $ {\delta{\cal V}_{xc}/ \delta
n(r)}$  become time-dependent. Together with the changing $A$, the electron 
in the unit cell drags its self-induced field with it, and the accelerations associated with 
these three terms in eq. (2) exactly cancel. The remaining ionic terms
then produce simple harmonic motion for $\phi$.  

\section{Numerical details}

The computational algorithm we employ is identical to the ones
we used for clusters and molecules, which was based on a method introduced
in nuclear physics\cite{fl78}. The Kohn-Sham operator is represented
on a real space grid as in ref. \cite{ch94}.  There are a number of technical 
details associated with the periodicity and with the gauge potential
that did not arise for the finite-system calculations.  In the new 
code, the potential
$V(r)$ is calculated by Fourier transformation of the Poisson
equation rather than a relaxation method.  This 
gives automatically the required periodicity to $V(r)$.    The wave 
functions $\phi_i$ represent Bloch states of the periodic lattice, and they 
are constructed with the corresponding periodic boundary conditions 
labeled by the Bloch momenta
$k$.  The periodic boundary condition on the Bloch wave function
$\phi_k(r + a) = \exp(ia k) \phi_k (r)$ is easily implemented in the
relaxation method used to find eigenstates.  In practice,
many Bloch states are needed to obtain smooth dielectric functions.
However, constructing the states takes much less time than for the
same number of electrons in a finite system, because the Bloch
states in a given band are automatically orthogonal.  

We use here the same energy density functional that we used previously
for finite systems.  Only the valence electrons are included explicitly; 
core electrons are treated by a pseudopotential\cite{tr91,kl82}.  The 
exchange-correlation energy of the electrons is calculated in the 
local density approximation following the prescription of ref. \cite{pe81}.  

The presence of a vector gauge potential requires a modification in
the pseudopotential calculation, as
indicated in the introduction.  In particular, the $A$-dependence of
the ${\cal V}_{ion}$ term in eq. (2) must be consistent with the
last term in eq. (3) in order to 
have the algorithm conserve energy.  
We implement the $A$-dependence of ${\cal V}_{ion}$ simply by
gauging the density matrix
directly, 
$${\cal V}_{ion,A}(\rho(r,r'))={\cal V}_{ion}(e^{i A (z-z')}
\rho(r,r'))$$
As in the finite systems calculations,
energy is conserved to a very high
accuracy with the algorithm \cite{fl78}, provided the time
step is less than the inverse energy span of the Kohn-Sham operator.

\section{Lithium}

In this section we demonstrate the feasibility of the method with
lithium as an example of a simple metal.  As other alkali metals, lithium
has a Fermi surface which is nearly spherical.  However, unlike
sodium and potassium, the effective mass of the electrons at the
Fermi surface is significantly enhanced over the free-Fermi gas
value ($m^* \approx 1.6 m_e$).  

The Kohn-Sham operator
is represented in coordinate space with a uniform spatial mesh.
The lattice spacing of the bcc unit cell of Li metal is $a= 3.49$~\AA, and we
take a mesh spacing of $\Delta x = 0.58$ to subdivide the cell
into a $6^3$ lattice of mesh points.  We use a time step of 
$\Delta t= 0.01$ eV$^{-1}$ which is sufficient to conserve energy
to 10$^{-4}$ eV over the time integration interval, $T=18$ eV$^{-1}$.

We sample the occupied states with a uniform mesh in momentum space.
With a finite set of occupied orbitals, the allowed excitation 
energies will be discrete, and the metallic behavior, $\epsilon(\omega)
\rightarrow \infty$, is only reached in the
limit of an infinitely dense momentum space lattice.  However, it
is our view that the TDLDA loses validity at long times when
other degrees of freedom can be excited.  This is the case for the
low frequency response of metals, where the imaginary part of the
dielectric responses is dominated by phonons and inelastic electron
scattering.

In Fig.~1 we show the normalized time-dependent polarization field,
$A_0^{-1} d A/ dt$ over the time interval $t=[0,18]$ eV$^{-1}$.  
The inset shows an expanded view of the initial response in the
interval [0,1] eV$^{-1}$.  The
dashed line is the comparison with the linear behavior deduced from
the sum rule, $A_0^{-1} d A/ dt= - 4 \pi e^2 N_e t / m \Omega$.  The
agreement shows that the local sum rule in nearly satisfied, despite
the fairly large optical effective mass.

In Fig.~2 we show the inverse dielectric function computed from
eq.~(4) for various meshes in the Brillouin zone.  We employ
$\eta=0.2$ eV to smooth the response in the Fourier transformation. 
We see that the response
becomes smoother, the more finely the Fermi sea is sampled.  With a $32^3$
lattice of Bloch states, we get results smooth enough to be compared
with measurement.

In Fig. 3  we show the real and imaginary parts of the inverse
dielectric function in the frequency interval $0-20$ eV/$\hbar$.  
The 
dashed lines show the empirical function from ref. \cite{crc}.
There is also another theoretical calculation in the literature,
\cite{br95}.  The agreement is quite good, especially considering
that the calculation is {\it ab initio} with a energy
density functional that much simpler than more recent ones.
The main feature in the absorptive response is the plasmon at
7 eV and its width.  The peak position is significantly downshift
from the naive plasmon frequency, $\omega = \sqrt{4 \pi e^2 n/m} \approx
8$ eV.  The width is associated with interband transitions and is
also well reproduced. 

\section{Diamond}

In this section we compute the dielectric response of a typical elemental
insulator, diamond.  The diamond lattice is represented in our
calculation by the primitive unit cell which contains 8 carbon atoms.
The four valence electrons of each carbon are calculated explicitly
while the core electrons are only treated implicitly by the pseudopotential.
We found in earlier studies of carbon structures that the Kohn-Sham Hamiltonian 
requires a mesh spacing $\Delta x =0.3$~\AA~to get orbital energies to an accuracy of
0.1 eV; in the calculation here we take a $12^3$ lattice in the unit
cell which implies $\Delta x =3.56/12\,\,{\rm \AA}= 0.297$~\AA.  With a 
smaller mesh spacing than for lithium, the span the Kohn-Sham operator is 
increased and
the time step $\Delta t $ must be reduced accordingly.  We use here
$\Delta t = 0.002$ eV$^{-1}$.

With a cubical unit cell and 8 carbon atoms, there are 
$8\cdot 4/2 = 16$ occupied bands\footnote{The bands are actually two-fold
degenerate because we have not exploited the symmetry that allows a smaller
unit cell with 4 carbons.}.  In each band we take
a lattice of up to $16^3$ points to represent the Bloch states.  

For an insulator, a reference point of the vector potential $A(t)$ should 
be irrelevant. Since all the points within Brillouin zone are occupied,
there is a static solution of eq.(2-3) for each constant vector potential.
However, in our real-space, real-time calculation, we found
this is violated due to the discrete representation in coordinate and
momentum spaces. The total energy does depend on $A$, and there appears 
a spurious low frequency mode in the vector potential $A(t)$ 
associated with the adiabatic evolution of the electronic wave 
function. Below that frequency, a spurious conduction feature 
appears in the response. 

These features are shown in the plots of the response in Fig.~4.
We see that the spurious adiabatic evolution gives an unphysical
plasmon at $\approx 1.2$ eV, which dominates the dielectric
response at lower frequencies. Though the amount of strength
associated with this spurious plasmon is very small, 0.007 electrons
out of the total of 32, it has a qualitative effect on the
response at very low frequency. The frequency and strength decrease
the finer the spatial mesh, showing that it is an artifact of the
discrete mesh representation of the coordinate space.

%Our initial calculations immediately revealed a small disadvantage of
%the formulation without explicit enforcement of the Pauli principle.
%Namely, the time-varying electron wave functions acquire amplitudes within
%as well as outside the bands.  
%The total inband response must be zero because there is only
%a single many-body state, with all inband transitions blocked by the
%Pauli principle.  In the usual methods, the Pauli principle is enforced
%explicitly by using the particle-hole representation.
%In the real-time formulation however, the inertness of fully occupied
%bands comes about by the cancelation of transitions involving different 
%electrons.  Since the cancelation is numerical, it is necessarily 
%imperfect.  There will thus appear a spurious conduction feature in
%the response.  The sum rule strength will be very small, but it will
%have a qualitative effect on the response at very low frequency.
%
%These features are shown in the plots of the response in Fig.~4.
%We see that the spurious in-band transition gives an unphysical 
%plasmon at $\approx 1.2$ eV, which dominates the dielectric
%response at lower frequencies.  However, the amount of strength associated 
%with this spurious plasmon is very small, 0.007 electrons out the total of 32.
%The strength also decreases the finer the mesh of Bloch states, showing
%that it is an artifact of the Brillouin zone sampling.  

To infer the
dielectric function near $\omega=0$, we apply the Kramers-Kronig relation to
the imaginary part of the response, but excluding the spurious plasmon peak.  
This gives the predicted dielectric function shown in Fig.~5.  The empirical
dielectric function is shown as the dashed line.  The agreement is good, as 
indeed was found solving the TDLDA equations by other methods\cite{ga97}, but
one can also see the effect of the well-known shortcoming of TDLDA, that
the predicted band gaps are too small\cite{hy86,go88}.  The 
theoretical absorption strength
become significant starting at about 5 eV excitation, while the empirical
absorption begins at around 7 eV.  Nevertheless, the dielectric constant
comes out in good agreement with the empirical\cite{ga97}, being within a 
percent of the empirical value of $\epsilon(0)=5.67$.

\section{Conclusions}
We see that the method not only works in principle, but produces
fairly accurate dielectric functions in the cases of a simple metal and a 
simple insulator.  In lithium, the theory describes the metallicity
as well as the interband transitions.  In diamond, there is a spurious
plasmon at low frequency 
%due to the Brillouin zone sampling. 
due to the discrete mesh representation in coordinate and momentum spaces.
However,
it can be easily dealt with and then the dielectric function
has an excellent quality except for a small band gap region. 
We find that two benefits of the real-space, real-time formulation of
the TDLDA in finite systems\cite{be00} are preserved in our implementation here.
The real-space method allows the Kohn-Sham
operator and electron-electron interactions to be evaluated efficiently
\cite{ch94}.  Computational efficiency is also gained by calculating
the response in real time in that all frequencies are calculated at
once.  Finally, the method requires much less storage than methods
using a particle-hole representation of the time-varying wave function.

\section{Acknowledgment}
We acknowledge discussions with E.K.U. Gross, R. Martin, J. Rehr, and R. Resta.  
This work was supported in part by the Department of Energy under
Grant DE-FG03-00-ER41132, by the DGES (PB98-0345) and by the JCyL
(VA28/99), and by the Ministry of Education, Science and Cultures (Japan), 
No. 11640372. JI and KY acknowledge the Institute of Solid State Physics,
University of Tokyo, and the Research Center for Nuclear Physics, Osaka
University for their use of supercomputers. AR acknowledges the hospitalty 
of the Institute for Nuclear Theory where this work was started and 
computer time provided by the Centre de Coputaci\'o i Comunicacions 
de Catalunya.

\begin{figure}
  \begin{center}
    \leavevmode
    \parbox{0.9\textwidth}
           {\psfig{file=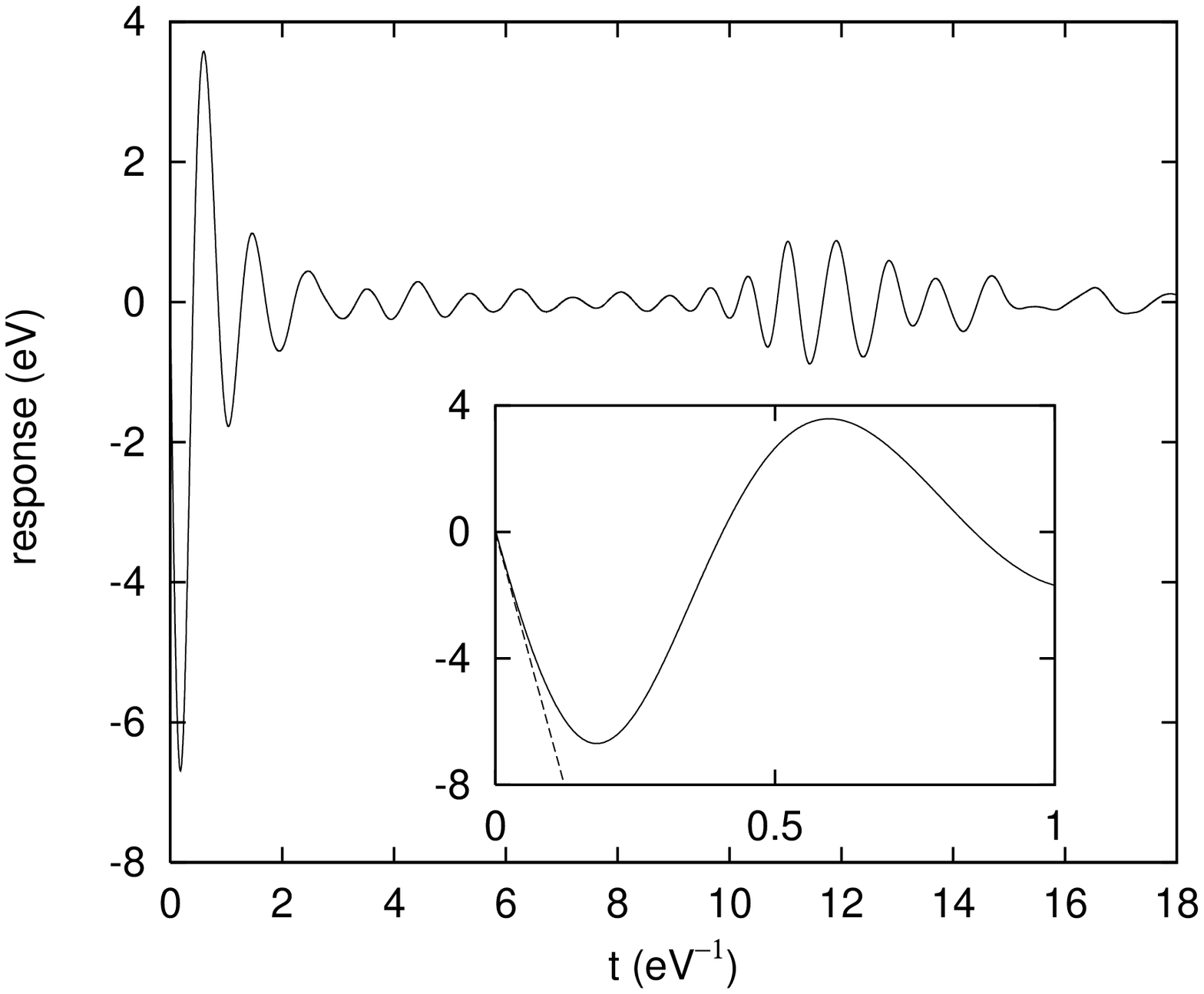,width=0.8\textwidth}}
  \end{center}
\protect\caption{
The induced polarization field $(1/A_0)dA/dt$ in lithium metal is
shown as a function of time (solid).  In this calculation, the occupied
states were represented by points on a $16^3$ mesh in $k$-space. The
dashed line show the early-time behavior required by the sum rule.
}
\end{figure}
\begin{figure}
  \begin{center}
    \leavevmode
    \parbox{0.9\textwidth}
           {\psfig{file=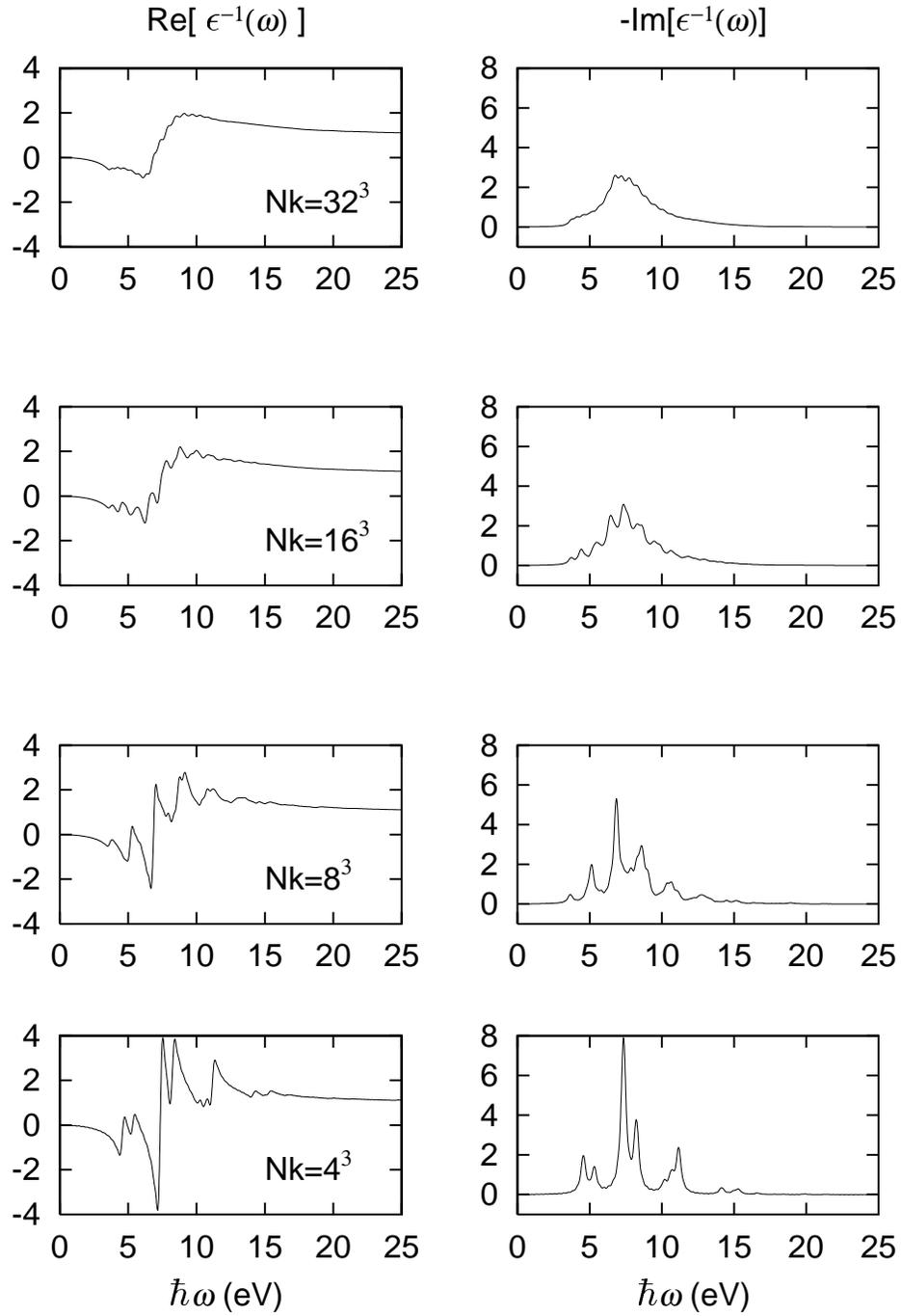,width=0.8\textwidth}}
  \end{center}
\protect\caption{
Real and imaginary parts of the inverse dielectric function
in lithium, shown for various meshes on the Brillouin zone.}
\end{figure}
\begin{figure}
  \begin{center}
    \leavevmode
    \parbox{0.9\textwidth}
           {\psfig{file=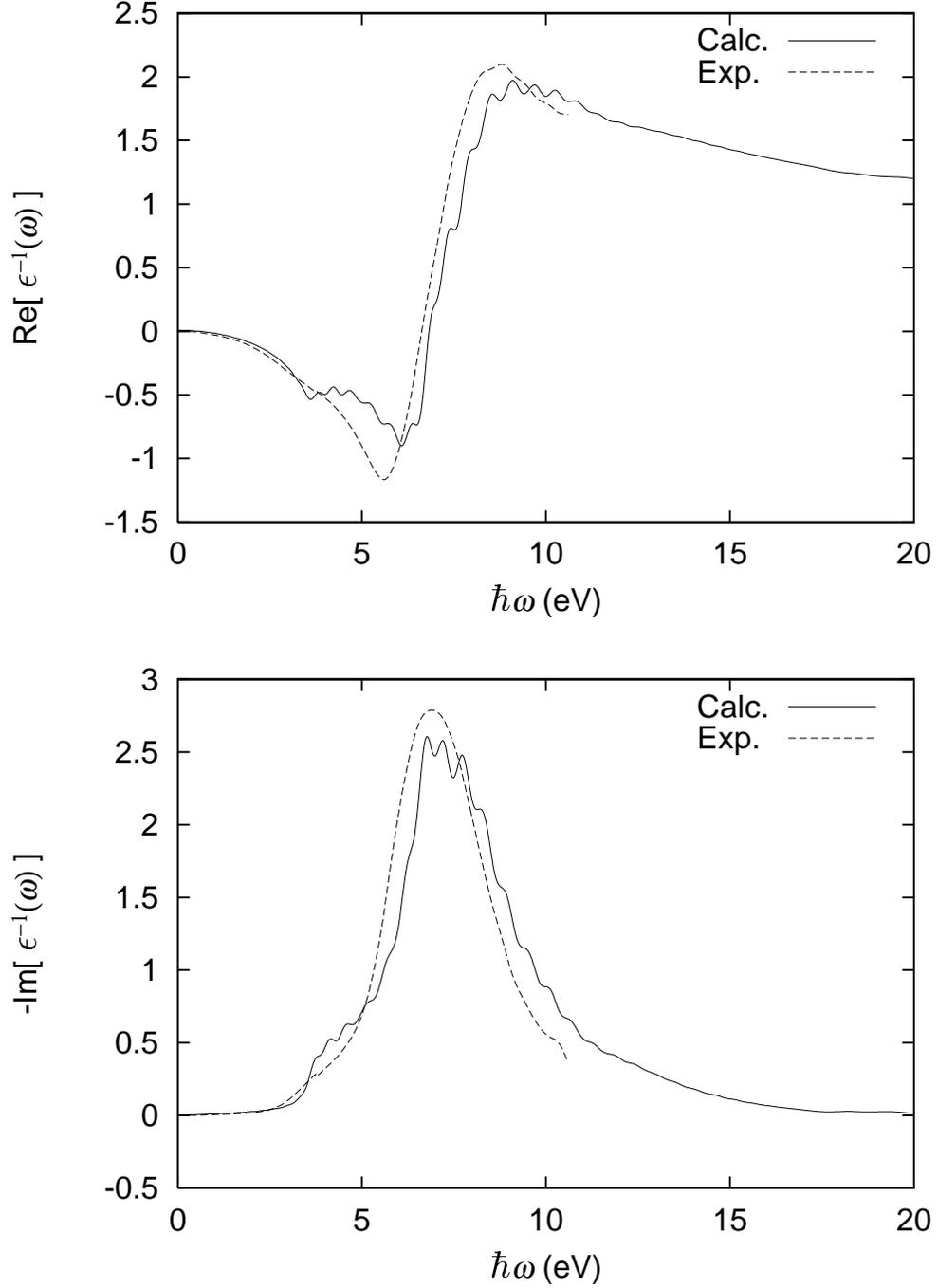,width=0.8\textwidth}}
  \end{center}
\protect\caption{
Real and imaginary parts of the response
$\epsilon^{-1}(\omega)$ as a function of frequency.  Here the
orbitals of the valence band were presented by Bloch states on
a $32^3$ mesh in $k$-space.
The empirical response from ref. \protect\cite{crc} is shown with 
the dashed lines.}
\end{figure}
\begin{figure}
  \begin{center}
    \leavevmode
    \parbox{0.9\textwidth}
           {\psfig{file=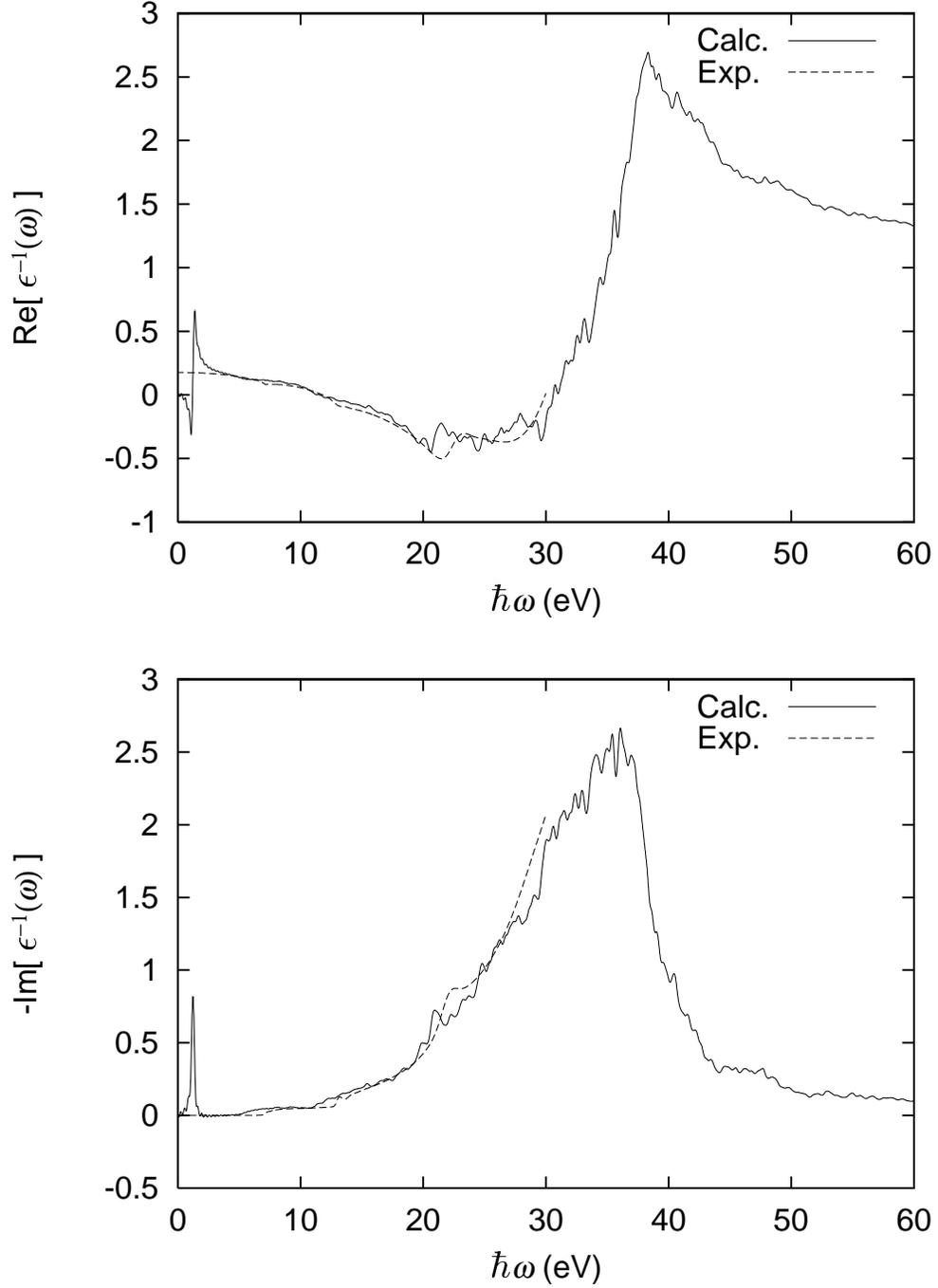,width=0.8\textwidth}}
  \end{center}
\protect\caption{
Real and imaginary parts of the response
$\epsilon^{-1}(\omega)$ for diamond.  Here the
orbitals of the valence band were represented by Bloch states on
a $16^3$ mesh in $k$-space.}
\end{figure}
\begin{figure}
  \begin{center}
    \leavevmode
    \parbox{0.9\textwidth}
           {\psfig{file=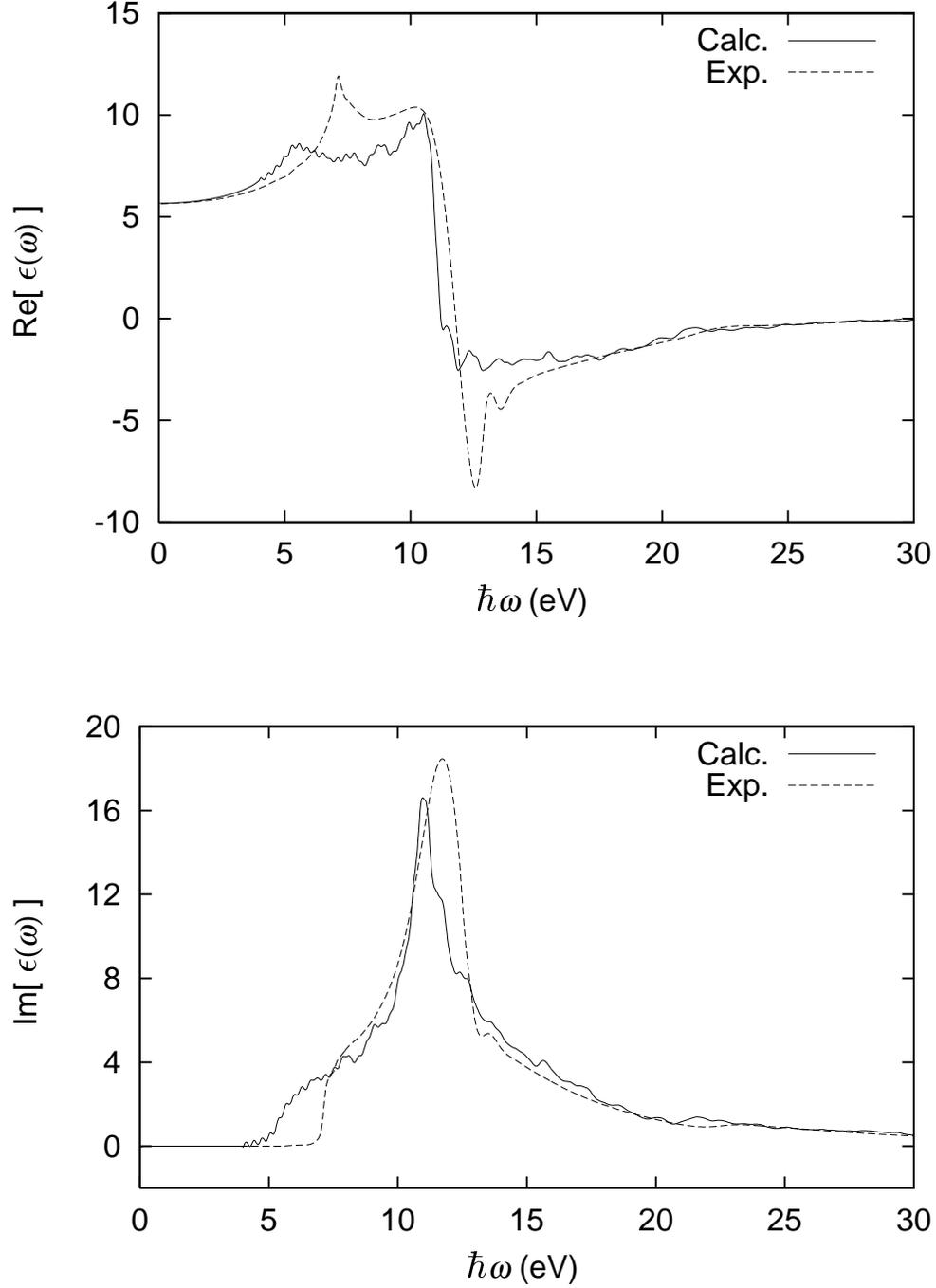,width=0.8\textwidth}}
  \end{center}
\protect\caption{
Real and imaginary parts of the dielectric function
$\epsilon(\omega)$ for diamond.  The spurious plasmon has been
excluded by using the Kramers-Kronig relation to determine the
real part of the dielectric function, integrating over the 
imaginary response from 4 eV.  The dashed curve shows the
experimental dielectric function, taken from ref.
\protect\cite{crc}.}
\end{figure}

\end{document}